# Integration of Behavioral Economic Models to Optimize ML performance and interpretability: a sandbox example.

Barranquero, R. (*), Gomez, Y. (**), Soria, E. (*) and Vila, J. (***)

* University of Valencia (IDAL) / ** DevStat / *** University of Valencia (ERI-CES and IDAL) and DevStat

## Abstract

This paper presents a sandbox example of how the integration of models borrowed from Behavioral Economic (specifically Protection-Motivation Theory) into ML algorithms (specifically Bayesian Networks) can improve the performance and interpretability of ML algorithms when applied to Behavioral Data.

The integration of Behavioral Economics knowledge to define the architecture of the Bayesian Network increases the accuracy of the predictions in 11 percentage points. Moreover, it simplifies the training process, making unnecessary training computational efforts to identify the optimal structure of the Bayesian Network. Finally, it improves the explicability of the algorithm, avoiding illogical relations among variables that are not supported by previous behavioral cybersecurity literature.

Although preliminary and limited to 0ne simple model trained with a small dataset, our results suggest that the integration of behavioral economics and complex ML models may open a promising strategy to improve the predictive power, training costs and explicability of complex ML models. This integration will contribute to solve the scientific issue of ML exhaustion problem and to create a new ML technology with relevant scientific, technological and market implications.

**Acknowledgments:** The authors thank the Spanish Ministry of Science, Innovation and Universities for financial support under project PID2019-110790RB-I00.

# 1. Introduction

Fueled with exponentially increasing data and computational power, ML is experiencing a dramatic development, in particular in its application to Behavioral Data. As currently approached, ML relays in a brute-force strategy based on Deep-Learning models with many parameters, reaching up to 17 billion, as in Google Turing-NLG (Rosset, 2020) and huge training datasets. Relevant voices in the field claim that the brute-force approach starts exhibiting diminishing marginal results and may be likely facing a critical wall (Marcus and Davis, 2019), in which is referred as the ML exhaustion problem. As an additional challenge, ML for Behavioral Data needs to deal with the inconsistency, instability, and noisiness of this type of data, which are significantly higher than in other data with more robust generation processes (mechanical, physical or biological data). Moreover, ML models with billions of parameters are difficult to explain. Lack of explicability of brute-force ML creates practical and ethical concerns that are limiting relevant market applications (e.g. as in autonomous driving or medical diagnoses).

After almost two centuries of human Behavior modelling based on axioms of rationality, Behavioral Economics (BE) has also paid attention to non-rational elements of human thinking. BE proposes models exhibiting a high capacity to explain empirical Behavioral insights that cannot be explained by rational-choice economics. The BE paradigm is now considered mainstream, as shown by the Nobel awarded by BE researchers since 2002 (Kahneman, Smith, Thaler, Dufflo, etc.) and the successful applications of the referred paradigm in policy-making and the industry.

This paper proposes a sandbox example of how the integration of Behavioral Economic Models (specifically Protection-Motivation Theory or PMT) into ML algorithms (specifically Bayesian Networks) can improve the performance and interpretability of ML algorithms when applied to Behavioral Cybersecurity Data. Specifically, we compare the results of a Bayesian Network whose underlying architecture is inferred from the training dataset with those of another Bayesian Network whose architecture is imposed ad-hoc by mimicking a validated behavioral economics model based in PMT. The results of the competition shows that the latter outperforms the former in accuracy of the predictions, efficiency of the training process and explicability of the algorithm.

## 2. Materials and methods

### 2.1. Protection-Motivation Theory

Protection-Motivation Theory (PMT) suggests that an individual's behavior is shaped by the evaluation of two cognitive appraisals: threat appraisal and coping appraisal (Figure 1). PMT was originally designed to explain engagement in protective actions in relation to health-related behaviors (Nasir et al, 2018). However, the theory has since been applied to the explanation of other protective actions, including uptake of insurance (Beck, 1984) and the adoption of secure online behaviors (Van Bavel et al., 2019).

PMT proposes that people protect themselves by making both a threat appraisal and a coping appraisal. The threat appraisal is dependent upon both the perceived severity of a threatening event (in this instance a cyberattack) and the perceived vulnerability to the event (i.e., the perceived probability of that event occurring). The coping appraisal reflects the perceived efficacy of the recommended protective behavior (cyberprotection in this case) and the individual's perceived self-efficacy (e.g., their ability to successfully use advanced cyberprotection measures). To explain: an individual considering whether to invest in cyberprotection may firstly weigh up the likelihood that they will receive a cyberattack of a particular severity against (a) the cost of taking out the advance protection measures (price to be paid, time, effort) and (b) how effective they believe that cyberprotection will be (response efficacy) and/or how much confidence they have in their own ability to put cyberprotection measures into place (self-efficacy).

*Figure 1. Protection-Motivation Theory*

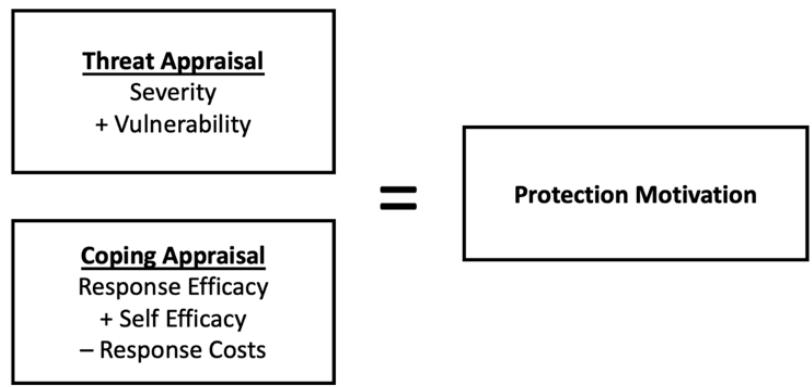

## 2.2. Bayesian Networks

A Bayesian network (BN) is a probabilistic graphical model composed of two different parts: on one hand is the underlying architecture (directed acyclic graph) that defines the relationship between variables and, on the other hand, the probabilities established between these variable (Koller and Friedman, 2009; Korb and Nicholson, 2011). The elements of a Bayesian network are as follows (Rusell and Norvig, 2009):

1. A set of variables (continuous or discrete) forming the network nodes.
2. A set of directed links that connect a pair of nodes. If there is a relationship with direction X →Y is said that X is the parent of Y.

The network fulfils the following facts:

- Each node $X_i$ is associated with a conditional probability function $P[Xi|Parents(Xi)]$ that takes as input a particular set of values for the node's parent variables and gives the probability of the variable represented by the node $X_i$.
- The graph has no directed cycles.

The knowledge is reflected by the relationships established in the graph nodes, and the conditional probabilities of the variables represented in each node.

## 2.3. The behavioral cybersecurity dataset

The data used to train the Bayesian Networks in this paper have been borrowed from an online behavioral economic experiment developed as part of the H2020 project CYBECO and presented in Branley-Bell et al. (2021). This experiment was designed and implemented to measure participants' cybersecurity decisions in a controlled situation and was mainly composed of two tasks: (i) Purchase decisions about cyberprotection measures products (cyber-security strategy), in particular the adoption or not of advanced security measure at a given costa and capable to reduce the risk of suffering a cyberattack; and (ii) online behavior whilst performing an online task. The instructions clearly explained all tasks and decisions to be made during the experiment and their implications. Figure 2 shows the experiment blueprint, which is described in detail in Branley-Bell et al. (2021).

*Figure 2. Blueprint of the economic behavioral experiment (Branley-Bell et al., 2021)*

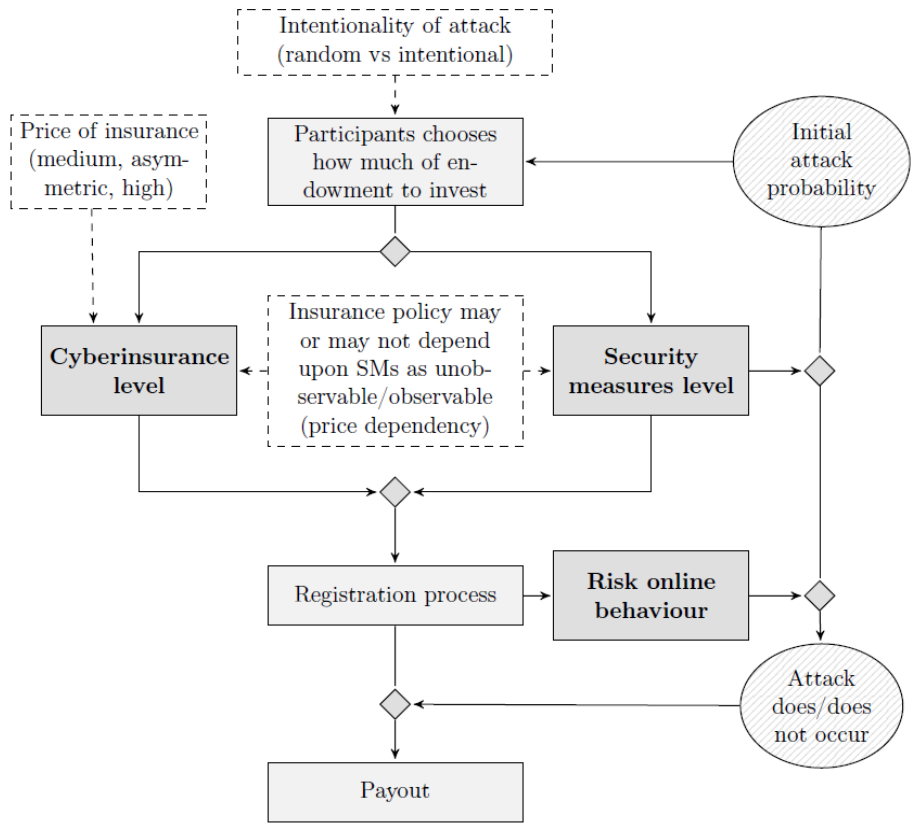

The final dataset used to train the Bayesian Networks consist of 3840 observations used to train the networl, and 960 more for evaluate their performance.

## 3. The proposed models

This example compares two Bayesian Networks (Pure and PMT-based Models) designed to predict the purchase of advanced cyberprotection measures by the subjects participating in the experiment described in subsection 2.1 using. The difference between both networks relies in their structure, i.e. the directed acyclic graph that defines the relationship between variables in the model. Meanwhile the architecture of the pure network is inferred from the training dataset using the Bayesian Augmented Tree algorithm (as described in detail in subsection 1.3), the architecture of the PMT-based network is prefixed according to pre-existing knowledge from the literature of behavioral cybersecurity.

### 3.1. *Pure ML model*

The TAN (Tree Augmented Network) is an extension of a Bayesian classifier in which each variable is allowed to have another parent outside the class node. The idea is to build a Bayesian network tree for all predictive variables and complete the model with a Naïve Bayes. TAN algorithm forms a tree with the predictive variables and then add edges to the class node; each predictive variable, $X_i$, can have up to two parents (Korb and Nicholson, 2011).

Figure 3 shows the TAN inferred from the 3840 observation of the training dataset.

*Figure 3. Bayesian Tree Augmented Model*

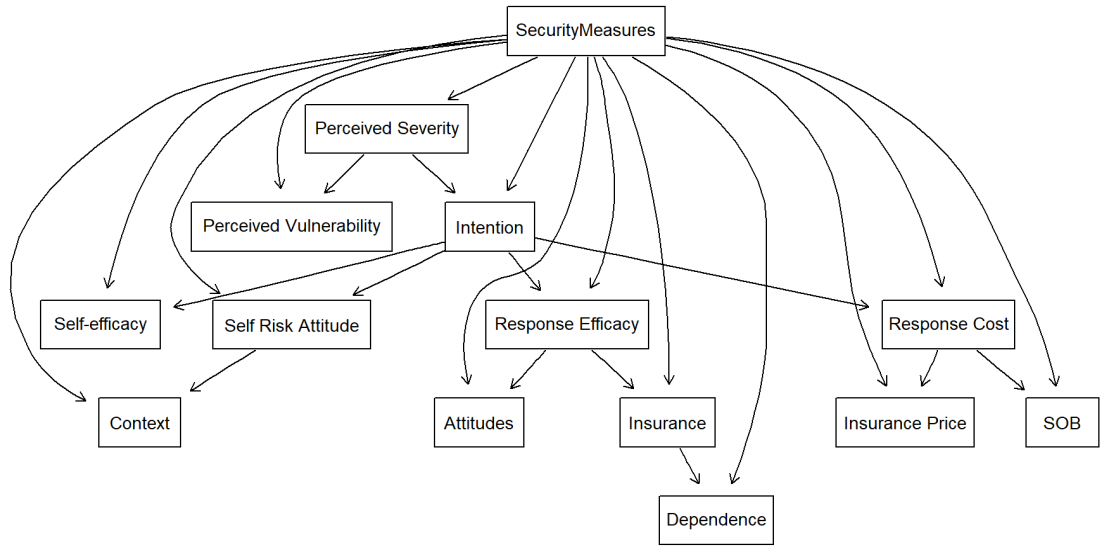

## 3.2. Behavioral PMT-based model.

Branley-Bell et al. (2021) apply PMT theory to develop a Structural Equation (SEM) predictive adoption model of cyberprotection and cyberinsurance. A graphical summary of this model is presented in Figure 4 (see the original paper for technical details).

*Figure 4. PMT-based Structural Equation Model of cyberprotection adoption with standardised coefficients. (Branley-Bell et al., 2021)*

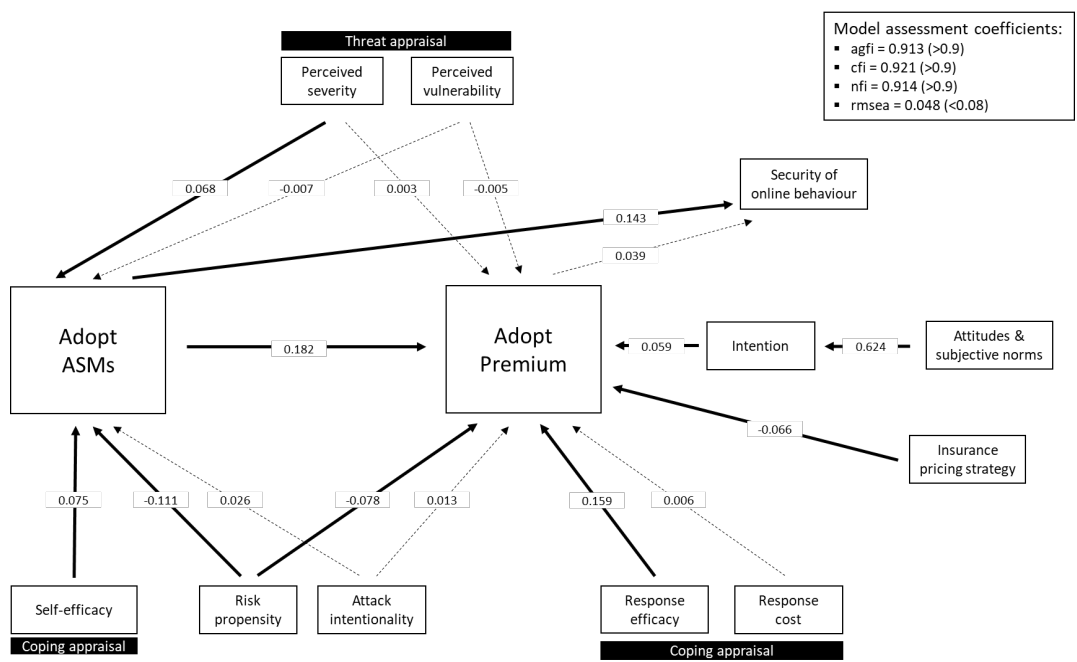

The architecture of our PMT-based Bayesian Network (Figure 5) has been predefined by mimicing the structural equations of the model in Figure 4. In this way, we simplify the training process (the algorithms needs not deal with the optimization of the network architecture) and integrate previous Behavioral Economics knowledge into our Machine Learning algorithm.

*Figure 5. Bayesian Network mimicking PMT model*

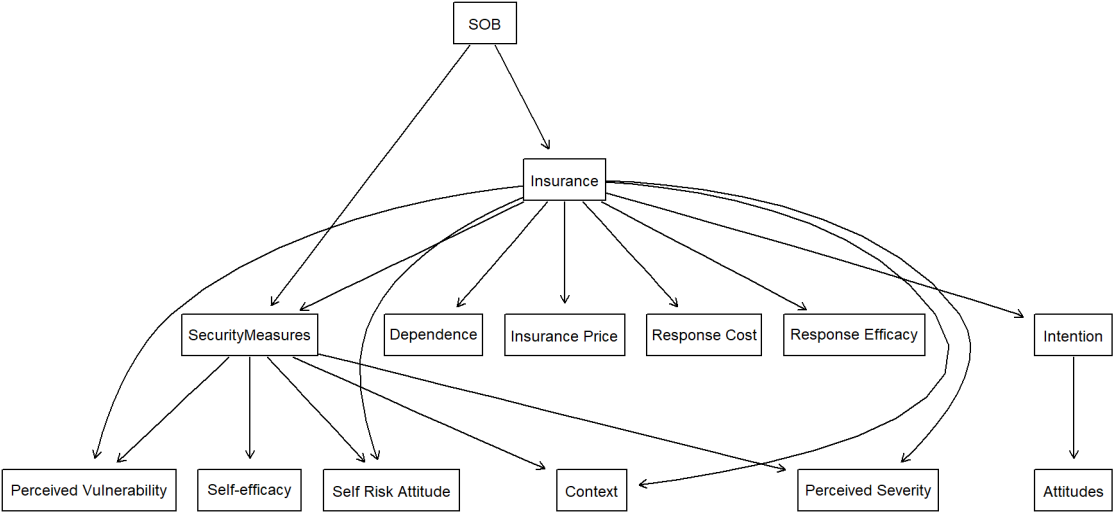

## 4. Results and discussion

Table 1 presents a comparison of the accuracy level achieved by both Bayesian Networks. Specifically, the table presents the percentage of right predictions of subjects purchasing the advanced security measures in the test dataset (960 observations).

*Table 1. Proportion of right predictions of subjects purchasing advanced security measures.*

| Model | % of right predictions |
|---|---|
| Pure ML model | 61% |
| PMT-based model | 72% |

As shown in this table, the integration of the PMT model to define the architecture of the Bayesian Network increases the accuracy of the network in 11 percentage points (from 61% to 72%) with respect to the pure ML model. It must be noticed that the integration of PMT model has improved the performance of the network even in our sandbox example, which considers only 14 variables and a small training dataset of size 3840.

The integration of behavioral economics insights into the structure of the Bayesian Network has two other relevant implications. First, establishing the architecture of the network ex-ante by mimicking the predictive cyberprotection adoption model in Branley-Bell et al. (2021) instead of inferring it from the training dataset, has simplified the training process. In particular, it makes unnecessary to run more complex algorithms to identify optimal network structure such the Bayesian Augmented Tree models. Therefore, the results of our example show that a restriction in the set of potential network architectures, when properly guided by behavioral economics knowledge, does not only compromise but improves the prediction capacity of the network, while reducing the computational efforts for model training.

Second, the integration of the PMT into the Bayesian Network improves significantly the explicability of the model. For instance, the Bayesian Augmented Tree Model, whose architecture has been inferred training the models with the cybersecurity dataset, considers influences among variables (such as a direct relation between perceived severity and intention) that are not supported by previous behavioral cybersecurity literature. Therefore, it is difficult to explain and to generate trust in a forecasting coming from this type of meaningless relations. On the other hand, all the elements in the PMT-based network are supported by the results of previous behavioral research and easy to explain to people for which may be affected by the outputs of the model. This type of explicability is a priority for the European Commission (which has even been incorporated this requirement into the General Data Protection Regulation, GDPR) and can help to cope with the lack of explicability of brute-force ML (which creates practical and ethical concerns that are limiting relevant market applications).

As final conclusion, although preliminary and limited to one simple model, our results suggest that the integration of behavioral economics and complex ML models may open a promising strategy to improve the predictive power, training costs and explicability of complex ML models. This integration will contribute to solve the scientific issue of ML exhaustion problem and to create a new ML technology with relevant market implications.